\documentclass[12pt]{article}
\usepackage{amssymb}
\usepackage{epsfig}

\newcommand{\bq}{\begin{equation}}
\newcommand{\eq}{\end{equation}}
\newcommand{\bqa}{\begin{eqnarray}}
\newcommand{\eqa}{\end{eqnarray}}
\newcommand{\ben}{\begin{enumerate}}
\newcommand{\een}{\end{enumerate}}
\newcommand{\bc}{\begin{center}}
\newcommand{\ec}{\end{center}}
\newcommand{\bqb}{\begin{eqnarray*}}
\newcommand{\eqb}{\end{eqnarray*}}

\def\pr#1#2#3{ Phys. Rev. ${\bf{#1}}$ (#2) #3}
\def\prl#1#2#3{ Phys. Rev. Lett. ${\bf{#1}}$ (#2) #3}
\def\pl#1#2#3{ Phys. Lett. ${\bf{#1}}$ (#2) #3}

\def\np#1#2#3{ Nucl. Phys. ${\bf{#1}}$ (#2) #3}

\global\nulldelimiterspace = 0pt

\begin{document}
\pagenumbering{arabic} \thispagestyle{empty}
\def\thefootnote{\fnsymbol{footnote}}
\setcounter{footnote}{1}

\vspace{4cm}

%---------------------titre ---------------------------------------

\begin{center}

{ \Large\bf Manifestations of Extra Dimensions \\
  \vspace{0.2cm} in a Neutrino
Telescope}
 \vspace{1.8cm}  \\
%-----------------------------------------------------------------
{\large A. Nicolaidis and D.T. Papadamou} \vspace {0.5cm}  \\

Department of Theoretical Physics, University of Thessaloniki,
GR-54006  Thessaloniki, Greece. \vspace{0.2cm} \\

\vspace {1.8cm}

{\bf Abstract}
\end{center}
Theories with large extra dimensions provide the possibility that
a flavor neutrino, localized in a 3+1 brane, can mix with a
singlet neutrino living in the bulk. This mixing leads to
unconventional patterns of neutrino matter oscillations and we
examine in details how these oscillations depend upon two
parameters: the brane-bulk coupling $\xi$ and the effective mass
$\mu$ of the flavor neutrino inside matter. We find that high
energy $(E \ge 50$ GeV) $\nu_\mu$ neutrinos, to be detected by
neutrino telescopes, can give signals of extra dimensions. With a
$1$ k$m^{3}$ neutrino telescope extra dimensions with radius down
to $1\mu m$ can be tested directly, while for smaller radius an
indirect evidence can be established.
\\

\vspace{0.5cm}
\def\thefootnote{\arabic{footnote}}
\setcounter{footnote}{0} \clearpage

\hoffset=-1.46truecm \voffset=-2.8truecm \textwidth 16cm
\textheight 23cm \setlength{\topmargin}{3cm}

{\bf Introduction}. The standard model (SM) of strong and
electroweak interactions has been extremely successful. It
provides a consistent theoretical framework within which we can
analyze and understand all available experimental data. Still we
know that it  cannot be regarded as the final theory. Any attempt
to include quantum gravity leads to a unified theory where two
disparate scales coexist, the electroweak scale ($M_W\sim 1$ TeV)
and the Planck scale ($M_{Pl}\sim 10^{19}$ GeV). Quantum radiative
corrections then, especially in the scalar sector (Higgs field) of
the theory, tend to mix the scales and without an incredible
amount of fine tuning, will always equalize them (hierarchy
problem).\par A novel approach has been suggested to alleviate the
hierarchy problem \cite{XR1}. Our four-dimensional world is
embedded in a higher dimensional space with D dimensions
($D=4+n$). While the SM fields are constrained to live on the
4-dimensional brane, gravity can freely propagate in the
higher-dimensional space (bulk). The fundamental scale $M_f$ of
gravity in D dimensions is related to the observed 4-dimensional
Planck scale $M_{Pl}$ by

\begin{eqnarray}
M^2_{Pl}= M^{2+n}_f V_n
\end{eqnarray}
where $V_n$ is the volume of the extra space. For a torus
configuration
\begin{eqnarray}
V_n=(2\pi)^nR_1R_2...R_n
\end{eqnarray}
with $R_i$ ($i=1,2,...n$) the radii  of extra dimensions. Then for
a sufficiently large volume $V_n$ the fundamental scale of gravity
$M_f$ can become as low as $M_W$. In this radical way the
hierarchy problem ceases to exist as such. \par Besides the
graviton, a  sterile neutrino can freely propagate in the extra
dimensions. Since the extra dimensions are compact, a bulk
neutrino will appear at the 4-dimensional brane as a Kaluza-Klein
(KK) tower, i.e. an infinite number of 4-dimensional spinors. The
Yukawa coupling of the standard lepton doublet, the Higgs scalar
and the right-handed bulk neutrino will provide a mixing between
the left handed neutrino of the standard model and the KK modes.
In the next part we present this mixing and we obtain the
eigenvectors and the eigenvalues of the mass matrix. In the third
part  we focus our attention into neutrino oscillations, inside
matter. Compared to the usual oscillations, novel features appear
since now we have a coupled system of infinite degrees of freedom.
The pattern of oscillations depends upon two parameters : the
coupling $\xi$ between the left-handed neutrino and the KK states
and the effective mass $\mu$ of the flavor neutrino inside matter.
We study in detail this dependence. In  the final part  we suggest
that high energy cosmic $\nu_{\mu}$ neutrinos can probe more
efficiently the bulk of the space. These neutrinos, after
traversing the Earth, will be detected at future neutrino
telescopes. We present detailed spectra of the comic $\nu_{\mu}$
as function of the energy and the incident angle. At the end we
present our conclusions .\\
\newpage

{\bf The Brane-Bulk Model}. The available gravity experiments
imply that any deviation from the Newton law should occur at
distances r less than 1 mm. Then, to satisfy equations (1)-(2), we
need more than two extra dimensions . To simplify our
investigation we consider a single large extra dimension, assuming
all other dimensions smaller and not affecting our analysis.\

There is already an extensive literature on neutrinos living in
extra dimensions \cite{XR2}-\cite{XR12a}. We start by considering
the action  for a massless bulk fermion $\Psi$ living in 5
dimensions
\begin{eqnarray}
S_{\Psi}=\int d^4x dy \bar{\Psi}\Gamma^{\mu} \partial_{\mu} \Psi
\hspace{1cm} M=0,1,...5
\end{eqnarray}
$\Psi$ is decomposed as

\begin{eqnarray}
\Psi=\left( \begin{array}{c}
      N_R \\
N_L \end{array} \right) \ \ \ \  \ \
\end{eqnarray}
each component admitting a Fourier expansion

\vspace{0.5cm}
\begin{eqnarray}
N(x,y)=\frac{1}{\sqrt{2 \pi R}} N_0 (x) + \sum_{n=1}^{\infty}
\frac{1}{\sqrt{\pi R}} \Biggl( N_n (x) \cos (\frac{ny}{R}) +
\hat{N}_n (x)\sin (\frac{ny}{R}) \Biggr)
\end{eqnarray}
Thus at the brane the bulk fermion appears as a KK tower. Since
the momentum in the exrtra compact dimension is quantized, each KK
mode appears as having a mass $n/R$ ($n=1,2,...$). Consider now
the coupling of the standard left-handed lepton doublet to the
bulk neutrino

\begin{eqnarray}
\frac{h}{\sqrt{M_f}}\hspace{0.1cm} \bar{L}HN_R \hspace{0.1cm}
\delta (y)
\end{eqnarray}
where H is the Higgs scalar doublet and h a dimensionless Yukawa
coupling. After the Higgs field develops a vacuum expectation
value $(v)$, we get the mass terms

\begin{eqnarray}
m\bar{\nu}_L ( N_{R0} + \sqrt{2}\hspace{0.15cm}
\sum_{n=1}^{\infty} N_{Rn})
\end{eqnarray}
with $m \sim h v M_f / M_{Pl}$. Notice that for $M_f \sim 1 TeV ,
h \sim 1$, we obtain $m \sim 10^{-4} eV$ \cite{XR2}-\cite{XR7}.
Also, the left handed zero mode $N_{L0}$ decouples from the
spectrum and remains massless. Denoting

\begin{eqnarray}
\Psi_R=\left( \begin{array}{c}
      N_{R0} \\
N_{Ri} \end{array} \right)  \ \ \ \  \ \ \Psi_L=\left(
\begin{array}{c}
      \nu_L \\
N_{Li} \end{array} \right) \ \ \ \ \  i=1,2,...
\end{eqnarray}
the mass term is $\bar{\Psi}_L M \Psi_R$ with the mass matrix M
given by  \cite{XR4}

\vspace{0.5cm}
\begin{eqnarray}
\begin{array}{cccc}
M= \bordermatrix { & m & \sqrt{2} m & \sqrt{2} m & \cdots
\\& 0 & 1/R & 0 & \cdots
\\ & 0 & 0 & 2/R & \cdots   \\ & \vdots & \vdots & \vdots & \vdots   \\}
\end{array}
\end{eqnarray}

\vspace{0.5cm} The evolution of the neutrino states is determined
by the Hamiltonian

\begin{eqnarray}
H = \frac{1}{2 E_{\nu}} M M^T
\end{eqnarray}
In terms of rescaled variables, $ H = X/(2 E_{\nu} R^2)$, where
the dimensionless matrix X becomes  (we keep k rows and columns)

\vspace{0.5cm}
\begin{eqnarray}
\begin{array}{cccccc}
X=\bordermatrix{& (2k+1) \xi^2 & \sqrt{2} \xi & 2 \sqrt{2} \xi & 3
\sqrt{2} \xi & \cdots & (k-1)\sqrt{2} \xi
\\ & \sqrt{2} \xi & 1 & 0 & 0 & \cdots & 0
\\ & 2 \sqrt{2} \xi & 0 & 4 & 0 & \cdots & 0 \\ & 3 \sqrt{2} \xi & 0 & 0 & 9 & \cdots & 0 \\
& \vdots & \vdots & \vdots & \vdots & \ddots & \vdots \\ &
(k-1)\sqrt{2} \xi & 0 & 0 & 0 & 0 & (k-1)^2 \\}
\end{array}
\end{eqnarray}
with $\xi = m R $. For a neutrino propagating inside matter the
upper-left element is replaced $(2k+1)\xi^2$ $\rightarrow$
$(2k+1)\xi^2+ \mu $, where $\mu$ is proportional to the effective
mass of the neutrino inside matter.For a $\nu_\mu$ neutrino
$\mu=\sqrt{2}E_\nu R^2 G_F N_n$ with $N_n$ the neutron density.
 The eigenvalues $\lambda^2_n$ of the matrix X satisfy the
 equation

\begin{eqnarray}
\Biggl[ \ \mu-\lambda^2+\xi^2 (\lambda \pi) \cot(\lambda \pi)\
\Biggr] \prod_{n=1}^{\infty} (n^2- \lambda^2)=0   \ \ \label{eigv}
\end{eqnarray}

The corresponding eigenvector to $\lambda^2_n$ is $ B_n = (e_{n0}
, e_{n1},e_{n2}, . . . )$ with \cite{XR4}

\begin{eqnarray}
e_{nk} = - \frac{k \sqrt{2} \xi}{(k^2- \lambda_n^2)}e_{n0} \ \ \ \
\  k = 1, 2, 3, . . .
\end{eqnarray}

\begin{eqnarray}
 e_{n0}^2 \Biggl(\frac{1}{2}+\frac{\pi^2
\xi^2}{2}+\frac{\mu}{2
\lambda_n^2}+\frac{(\lambda_n-\frac{\mu}{\lambda_n})^2}{2 \xi^2}
\Biggr) = 1   \ \ \label{eno}
\end{eqnarray}

\vspace{1.0cm}

\begin{figure}[p]

\begin{center}
\mbox{ \epsfig{file=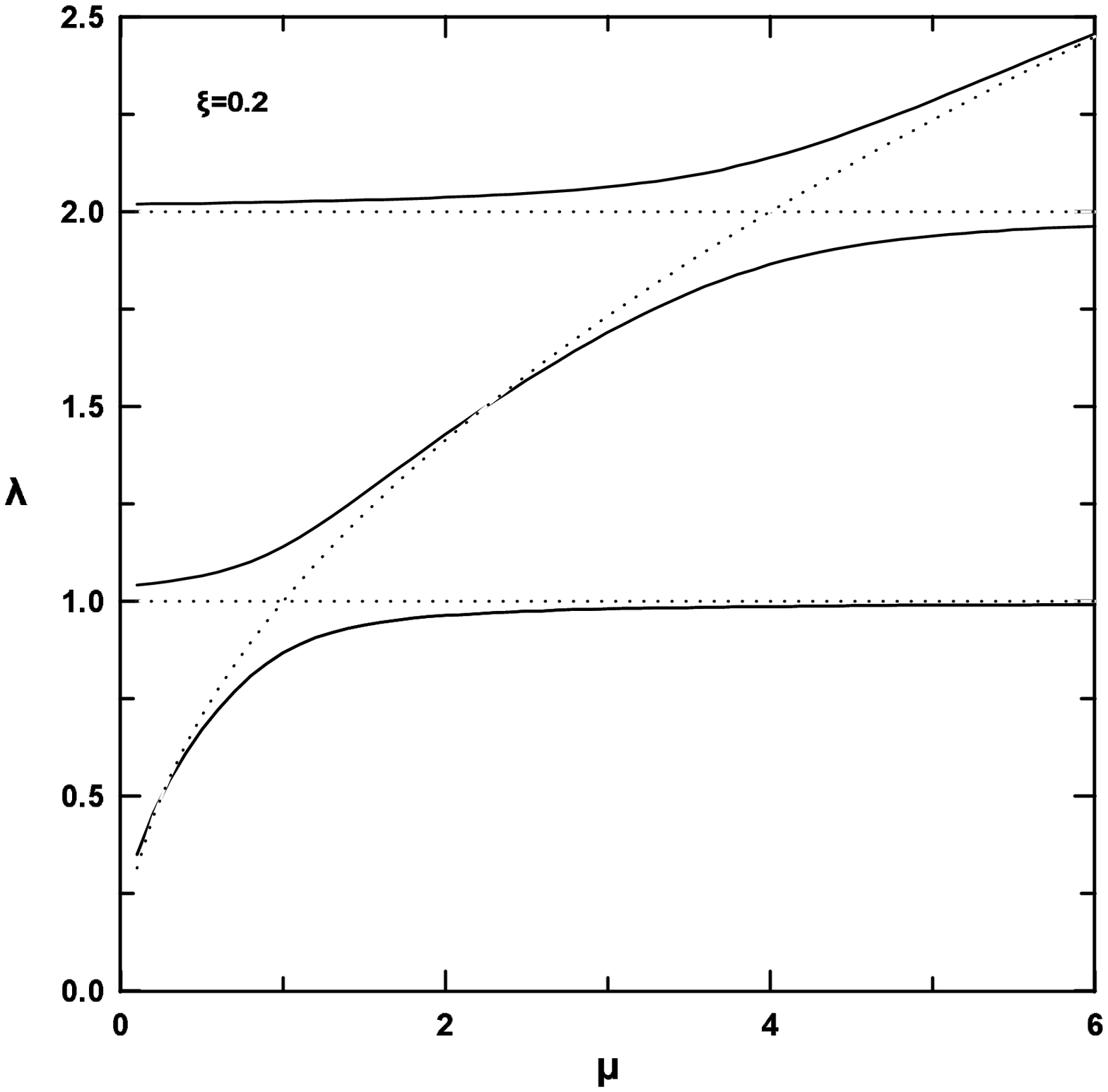,height=12cm}} \vspace{2.0cm}
\end{center}  \caption[1]{Eigenvalues $\lambda$ as a function of $\mu$. Dashed lines for
$\xi=0$ (no coupling among the SM neutrino and the KK states) and
the solid lines for $\xi=0.2$ } \label{fig1}
%\end{center}
\end{figure}

\begin{figure}[p]

\begin{center}
\mbox{ \epsfig{file=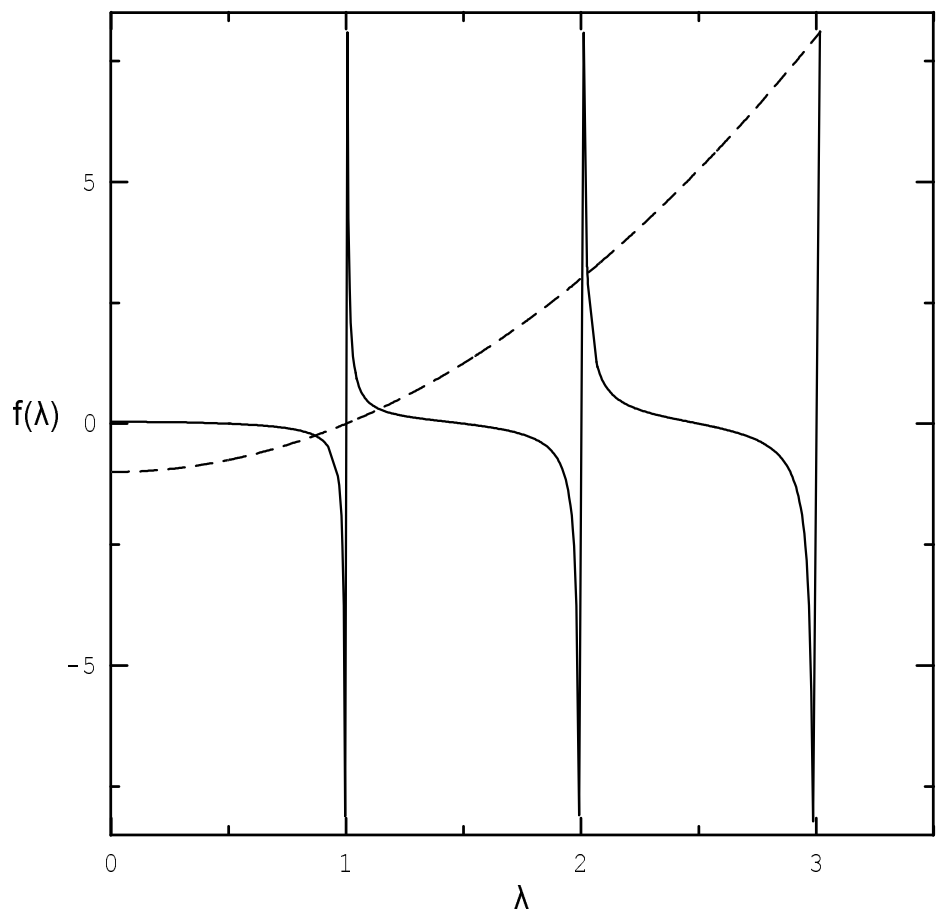,height=12cm}} \vspace{2.0cm}
\end{center}   \caption[1]{ Dash line for $
f(\lambda)=\lambda^2-\mu$ and solid line for $ f(\lambda)= \xi^2 (
\lambda \pi ) \cot( \lambda \pi )$ with $\xi=0.2$} \label{fig2}
%\end{center}
\end{figure}

{\bf Neutrino Mixing}. Before presenting the numerical solution to
equation (\ref{eigv}) it is useful to mention some qualitative
features

   i) small $\xi$ \\ For $\xi=0$ there is no coupling between the
SM neutrino and the KK states. The left-handed neutrino remains
massless, $ \lambda_0=0 $, while for a neutrino propagation inside
matter $ \lambda_0 =\sqrt{\mu} $. The other eigenvalues are
$\lambda_n = n \ \ ( n =1, 2, 3, ...)$. As $\xi$ starts to rise $
\lambda_n \simeq n + \xi^2 / (n \pi) $ (for large n)

  ii) large $\xi$ \\ For large $\xi, \lambda_n$ tend to $ n+1/2
$. An expansion for large $\xi$ and large n provides

\begin{eqnarray}
\lambda_n\simeq (n+\frac{1}{2}) (1-\frac{1}{\xi^2 \pi^2})
\end{eqnarray}

iii) resonance structure \\
 The amount of admixture of the n-th eigenstate $B_n$ to the
 neutrino state $\nu$ is determined by  $|e_{n0}|^2$ and it is
 becoming maximal when the resonance condition
\begin{eqnarray}
\mu_R = \lambda_n^2
\end{eqnarray}
 holds (see equation \ref{eno}). Since $\lambda_n \simeq n $, the
 resonance condition is simplified to $ \mu \simeq n^2 \ \ \ (n=1,
 2, 3, ...)$. For small $\xi $ values the resonance is narrow,
 while for large $\xi$ values the resonance is broad and more
 than one state might be involved. Recall that for a neutral
 medium

\begin{eqnarray}
\mu = \frac{G_F R^2 \rho E_{\nu} }{\sqrt{2} M_N} \ \ \label{mu}
\end{eqnarray}
with $\rho$ the density of the medium and $M_N$ the nucleon mass.
The requirement $\mu\simeq1$  (strong oscillation phenomena) leads
to a relationship between the model parameters (R), the medium
properties ($\rho$) and the neutrino energy ($E_{\nu}$)
\begin{eqnarray}
( \frac{\rho}{10 \ \frac{gr}{cm^3}}) \ (\frac{E_{\nu}}{100 \ GeV})
\ (\frac{R}{1 \ \mu m})^2 \  \simeq 1 \ \ \label{muo}
\end{eqnarray}

   iv) Level repulsion \\ When there is no
coupling among the SM neutrino and the KK states ($\xi=0$), the
eigenvalues cross each other (see dashed lines in figure
\ref{fig1} ). When the off diagonal elements become non-zero $(
\xi\neq
 0)$, the well-known phenomenon of level repulsion appears : as
 the lower eigenvalue increases by increasing $\mu$, by approaching
 the resonance region  $( \mu \simeq 1 )$ levels off and reaches
 asymptotically the value $1$, thus avoiding crossing the next
 eigenvalue. Similarly the higher eigenvalue starts at $ \lambda
 \simeq1 $, increases as $\mu$ increases and when encounters the
 second resonance $( \mu \simeq 4 )$ levels off at the value 2.
 (see the figure \ref{fig1} )

    v) graphical solution \\
 The eigenvalues $ \lambda_n^2$ can be found graphically
 as the intersection points of the two curves  (see the figure
\ref{fig2} )

\begin{eqnarray}
 y_1 =
\lambda^2 - \mu \end{eqnarray}

\begin{eqnarray}
y_2 = \xi^2 ( \lambda \pi ) \cot( \lambda \pi )
\end{eqnarray}

Most of the qualitative figures described above can been seen by
inspecting the figure \ref{fig2}.

\vspace{1.0cm}

{\bf Oscillating Neutrinos and Neutrino Telescopes}. From the
previous analysis it is clear that for neutrinos traveling inside
the Earth, strong oscillation phenomena occur whenever the
condition (\ref{muo}) is satisfied. Thus for $ R\sim 1\mu m $, we
need neutrinos of energy $E\sim 100 $ GeV, while a smaller radius
is probed by even higher energy neutrinos. It is clear that
neutrino telescopes \cite{NT} which are able to detect $ \nu_{\mu}
$ neutrinos from $\sim 20$ GeV up to $ \sim 50 $ TeV are best
suited for this study. At this early stage of investigation we
adopt a simple model where the $\nu_{\mu}$ flavor state mixes only
with a bulk neutrino. We expect this simple model will indicate
the salient features of the more complicated physical phenomenon.
\par

\begin{figure}[p]

\begin{center}
\mbox{ \epsfig{file=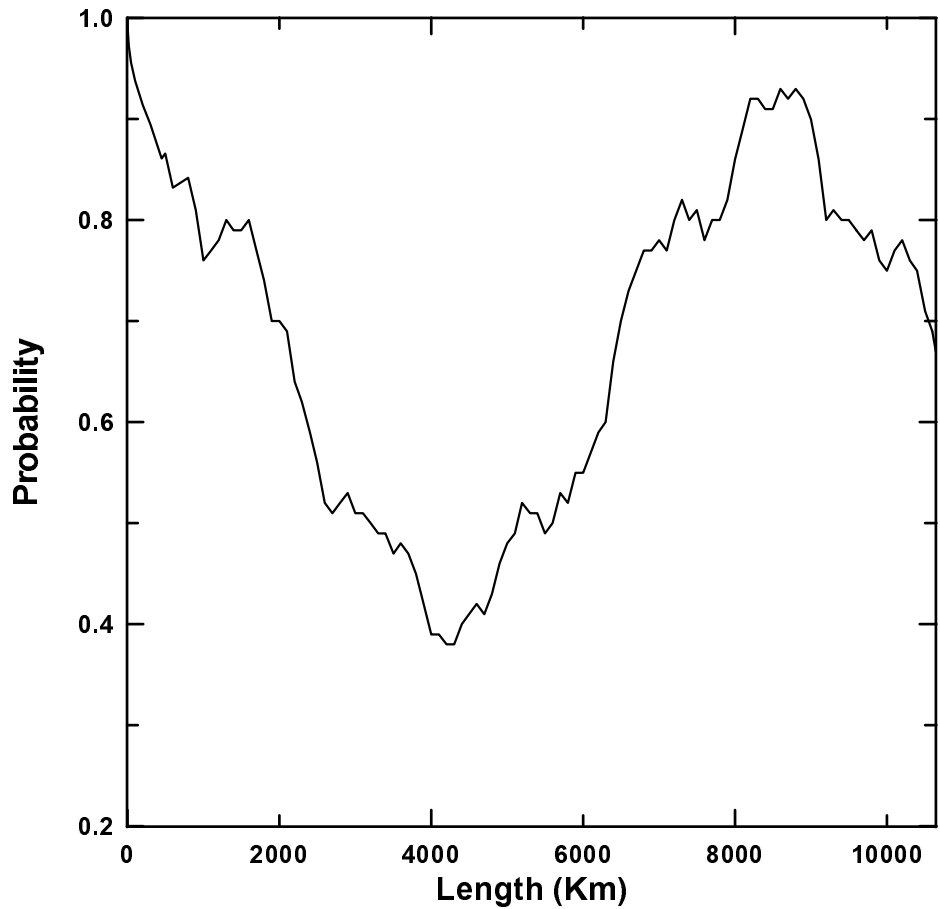,height=12cm}} \vspace{2.0cm}
\end{center}  \caption[1]{ The probability $P(\nu_\mu \rightarrow
\nu_\mu )$ for a neutrino of energy $100 GeV $ traversing a medium
with  density $\rho=4.4 \ gr/cm^3$ (close to the density of the
Earth's mantle) as a function of distance. } \label{fig3}
%\end{center}
\end{figure}

\begin{figure}[p]

\begin{center}
\mbox{ \epsfig{file=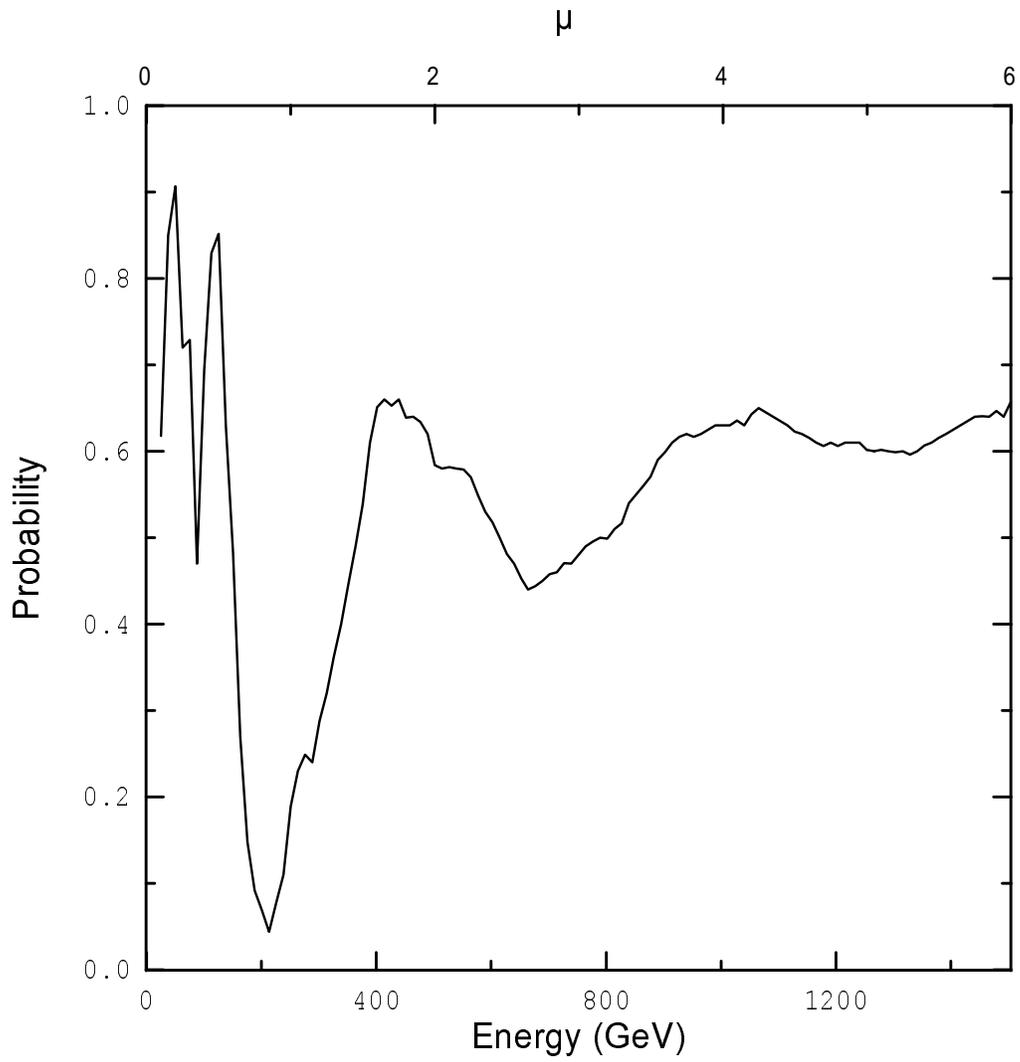,height=14cm}} \vspace{2.0cm}
\end{center}   \caption[1]{The probability $P(\nu_\mu \rightarrow
\nu_\mu )$ for a neutrino  traversing 10649 Km through the mantle
of Earth ( with density $\rho=4.4 \ gr/cm^3$), as a function of
energy.} \label{fig4}
%\end{center}
\end{figure}

Once a $\nu_{\mu}$ flavor state is produced, its time evolution
will be governed by the equation

\begin{eqnarray}
|\nu (t)> = \ \sum_{n} \  e_{n0}^* \exp ( \frac{-i \lambda_n^2
t}{2 E_\nu R^2}) \ \ |B_n>
\end{eqnarray}
where the summation over n is carried  up to $n_{max}=[E_\nu R]$
The amplitude for the transition $ \nu_\mu\rightarrow \nu_\mu$ is

\begin{eqnarray}
A = \ \sum_{n} \  |e_{n0}|^2 \exp ( \frac{-i \lambda_n^2 t}{2
E_\nu R^2})
\end{eqnarray}
Figure \ref{fig3} shows the probability $P(\nu_\mu \rightarrow
\nu_\mu )$ for a neutrino traversing a medium of constant density,
as a function of distance. The value we used for the density of
the medium corresponds to the density of the Earth's mantle. We
used also $R=1\mu m$. A clear dip develops at the appropriate
length scale. The probability $P(\nu_\mu \rightarrow \nu_\mu )$ as
a function of energy is presented in figure \ref{fig4} . Notice
that an impressive resonance structure appears at the energy which
corresponds to $\mu \simeq 1$ and all $\nu_\mu$ disappear. At the
resonance, reading the value of the energy and using equation
(\ref{muo}) we can extract the value of the radius R.
\par
 It is useful to
evaluate the transition probability for a neutrino traversing the
Earth. For the Earth's density we adopt a model where the core is
a sphere with constant density $11 \ gr/cm^3$ while the mantle,
surrounding the core has constant density $4.4 \ gr/cm^3$
\cite{nic}. We consider a neutrino which for time $t_1$ propagates
inside the mantle, for time $t_2$ crosses the core
 and it takes time $t_3$ within
the mantle again . The amplitude $A(\nu_\mu \rightarrow \nu_\mu )$
is given by

\begin{eqnarray}
A = \ \sum_{n,l,r} \  e_{n0}^* \exp ( \frac{-i \lambda_n^2 t_1}{2
E_\nu R^2}) \ D_{nl} \ \exp ( \frac{-i \tilde{\lambda}_l^2 t_2}{2
E_\nu R^2}) \ D_{lr}^{-1} \ \exp ( \frac{-i \lambda_r^2 t_3}{2
E_\nu R^2}) \ e_{r0}
\end{eqnarray}
where tilded (untilded) parameters correspond to core (mantle) and
D is the rotation matrix from one set of eigenvectors to the other

\begin{eqnarray}
D_{nl} = \ \sum_{k} \  e_{nk}  \tilde{e}_{lk}^*
\end{eqnarray}

\begin{figure}[p]

\begin{center}
\mbox{ \epsfig{file=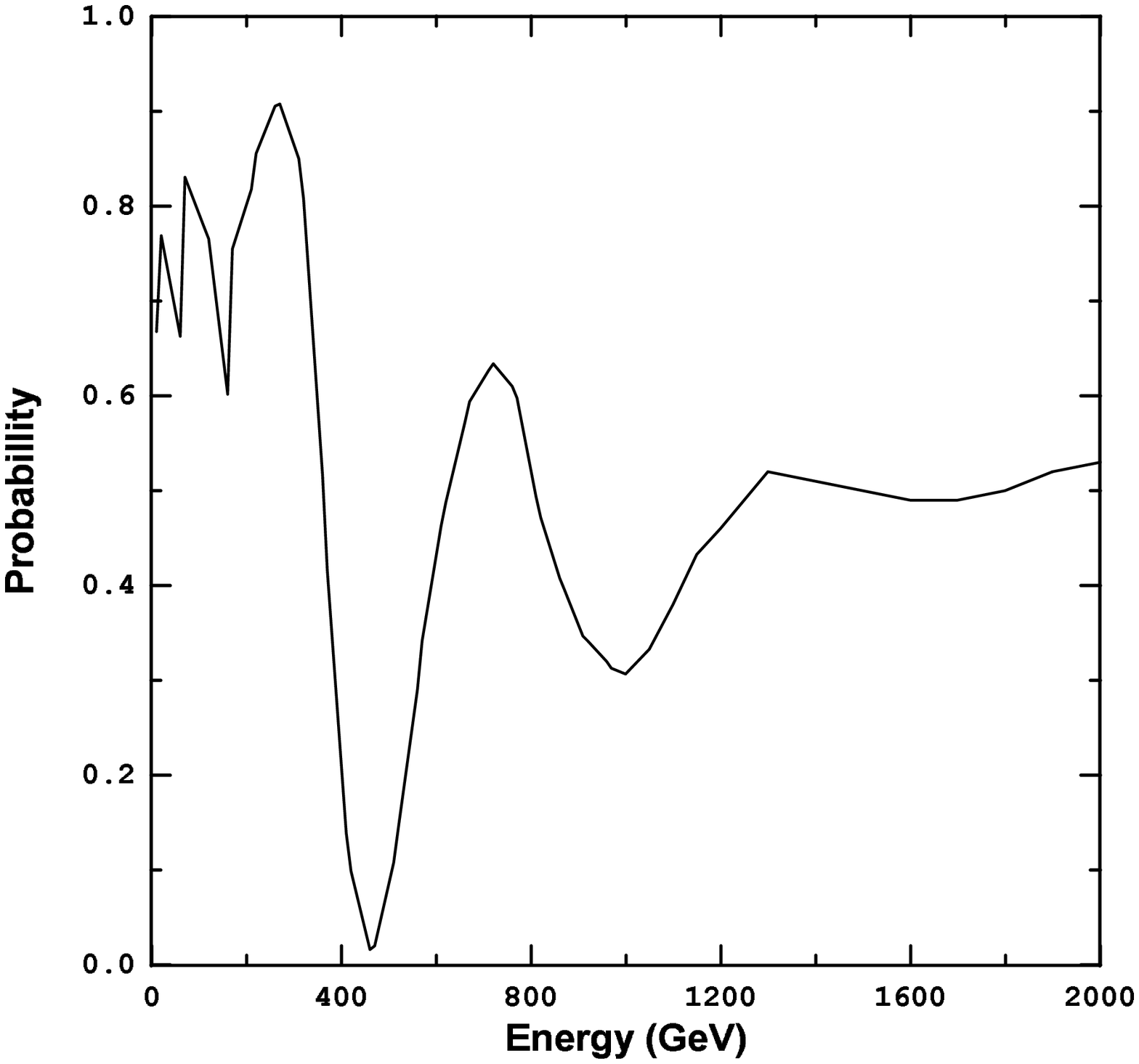,height=14cm}} \vspace{2.0cm}
\end{center}   \caption[1]{The probability $P(\nu_\mu \rightarrow
\nu_\mu )$ for a neutrino  going through the center of the Earth ,
as a function of energy.} \label{fig5}
%\end{center}
\end{figure}

The probability $P(\nu_\mu \rightarrow \nu_\mu )$ for a neutrino
going through the center of the Earth, as a function of the energy
is shown in figure \ref{fig5}. The resonance dip is now
displaced.\par How efficiently can a neutrino telescope unravel
the underlying dynamics? Without entering into technical details
and specifications of a neutrino telescope, we can recall that by
determining the direction of the produced muon, we also fix  the
direction of the parent muon neutrino. Thus we can have a zenith
coverage of incident neutrino from $\theta=0$ (neutrinos going
through the center of Earth) to $\theta\simeq\pi/2$ (neutrinos
just scratching the Earth, before being detected). In the absence
of any oscillations the neutrino flux is  $\theta$-independent. If
oscillations are present, by varying $\theta$ we vary the distance
x traveled by $\nu_{\mu}$ (x varies continuously from $10$ km for
$\theta=\pi/2$ to $13000$ km for $\theta=0$). In our case the
$\theta\simeq\pi/2$ neutrino flux remains undistorted by matter
effects and represents  therefore the initial incident neutrino
flux. The deduced ratio $I(\theta)/I(\theta=\pi/2)$ $(I(\theta)$
is the infered intensity of neutrino flux at angle $\theta$)
reflects then the probability for a muon neutrino to oscillate.The
energy of the incident neutrino is determined for the contained
events \cite{smoot}. Thus we have a set of events, where both the
distance x and the energy of the neutrino $E_{\nu}$ are
determined, allowing to extract the theoretical parameters ($\xi$
and R). Neutrino induced events with energy up to $300 - 400$ GeV
are expected to be contained within a $1$ k$m^{3}$ volume and
therefore we deduce that $1$ k$m^{3}$ neutrino telescope can
provide direct evidence for an extra dimension with $R\geq 1\mu
m$. For smaller R, we expect the resonance structure to shift to
higher energies and thus the resonance dip will not be directly
seen. Still we may have an indirect evidence for strong
oscillations by examining the ratio $r=E_{s}/E_{p}$, with $E_{s}
(E_{p})$ the number of stopping (passing) muons \cite{lipari}. If
at $\theta=0$ the number of highly energetic $\nu_{\mu}$   $(E_\nu
\geq 400$ GeV) is depleted, then we expect the ratio r to increase
for $\theta=0$ compared to the same ratio for $\theta\simeq\pi/2$.
\par Another important scale in the neutrino oscillations is the
resonance oscillation length $L_{R}$. A significant depletion of
$\nu_{\mu}$ occurs whenever the distance traveled x is equal to
the  oscillation  length L, defined by
\begin{eqnarray}
L = 2 \pi \frac{E_\nu R^2  }{\lambda^{2}_{1}-\lambda^{2}_{0}} \ \
\label{L}
\end{eqnarray}

At the resonance , $\mu\simeq1$, L becomes
\begin{eqnarray}
L_{R} = \frac{2\sqrt{2} \ \pi \ M_{N}}{G_{F} \ \rho \
(\lambda^{2}_{1}-\lambda^{2}_{0})} \ \ \label{Lr}
\end{eqnarray}

Whenever both equations (\ref{muo}) and (\ref{Lr}) are satisfied,
we observe the maximum depletion of $\nu_{\mu}$
($P(\nu_{\mu}\rightarrow \nu_{\mu})\simeq 0$). The eigenvalues
difference $\lambda^{2}_{1}-\lambda^{2}_{0}$ depends upon $\xi$.
For small $\xi$ values the resonance is narrow and the difference
$\lambda^{2}_{1}-\lambda^{2}_{0}$ is small. The corresponding
$L_{R}$ is big, much bigger than the Earth's diameter. For example
for $\xi=0.05$ and with $\rho = 10 gr/cm^{3}$ we find $L_{R}=
23571$ km. In that case strong oscillations cannot be observed.
For $\xi = 2.2$ (the value we opted), we find $L_{R} = 6000$ km,
smaller than Earth's size, and strong oscillations are visible.
 \vspace{1.0cm}

{\bf Conclusions}. We analyzed the situation where the left-handed
neutrino $\nu_{\mu}$ of the SM mixes with a right handed neutrino
experiencing a large extra space dimension. This mixing induces
neutrino oscillations, even if the neutrino itself is massless. We
encountered novel features compared to the standard oscillations,
since the mixing involves an infinity of KK states. A rich
resonance structure appears accompanied by many spikes. We
presented in detail the first resonance ($\mu \simeq 1$). A
similar study can be realized for the second resonance ($\mu
\simeq 4$). We emphasized that $1$ k$m^{3}$ neutrino telescope may
be used as an instrument to explore the bulk space and establish
unambiguously the existence of extra dimensions with $R\geq 1\mu
m$. We didn't address the issues related to low energy neutrino
data (solar neutrinos, atmospheric neutrinos), since for the R
values we used, there is no significant mixing of low energy
neutrinos with the bulk states. After all, the recent neutrino
data (SNO  \cite{sno}) disfavor the presence of a sterile
neutrino. Altogether our analysis shows that neutrino mixing in
models with large extra dimensions is an interesting phenomenon
and deserves further study.
\\
\newpage

%\newpage

%%%%%%%%%%%%%%%%%%%%%%%%%%%%%%%%%%%%%%%%%%%%%%%%%%%%%%%%%%%%%%%%%%%%%%%

\clearpage
\newpage

\clearpage
\newpage

\clearpage
\newpage

\clearpage
\newpage

\clearpage
\newpage

\clearpage
\newpage

\clearpage
\newpage


\begin{thebibliography}{99}


\bibitem{XR1} I. Antoniadis \pl{B246}{1990}{377};
N. Arkani-Hamed, S. Dimopoulos and G. Dvali ,
\pl{B429}{1998}{263}; I. Antoniadis, N. Arkani-Hamed, S.
Dimopoulos and G. Dvali , \pl{B436}{1998}{257}.
%
\bibitem{XR2} N. Arkani-Hamed, S. Dimopoulos, G.Dvali and J. March-Russsell
hep-ph/9811448.
%
\bibitem{XR3}K. Dienes, E. Dudas and T. Ghergetta \np{B557}{1999}{25} .
%

\bibitem{XR4} R. Barbieri, P. Creminelli and A. Strumia,
\np{B585}{2000}{28}.

%
\bibitem{XR5} R. Mohaparta, S. Nandi and Perez-Lorenzana,
\pl{B466}{1999}{115}.

%
\bibitem{XR6} A. Lukas, P. Ramond, A. Romanino and G Ross, \pl{B495}{2000}{136}.
%
\bibitem{XR7} G. Dvali and A. Smirnov, \np{B563}{1999}{63}.
%
\bibitem{XR8} K. Dienes and I Sarcevic, \pl{B500}{2001}{133}.

%
\bibitem{XR9} A. Das and O. Kong, \pl{B470}{1999}{149}.

%
\bibitem{XR10} D. Caldwell, R. Mohaparta and S. Yellin
hep-ph/0102279.
%
\bibitem{XR11} N. Cosme, J. Frère, Y. Gouverneur, F. Ling,
D. Monderen and V. Van Elewyck, \pr{D63}{2001}{113018}.
%
\bibitem{XR12} C.S. Lam and J.N Ng hep-ph/0104129.
%
\bibitem{XR12a} A.Gouvea, G. Giudice, A. Strumia, K. Tobe, hep-ph/0107156.
%
\bibitem{NT} The AMANDA Collaboration Astropart, Phys {\bf 13}, 1
(2000) .\\ The NESTOR Collaboration  \np{Proc. Suppl \
87}{2000}{448}.\\ The ANTARES Collaboration  \np{Proc. Suppl \
81}{2000}{174}.
%
\bibitem{nic} A. Nicolaidis \pl{B200}{1988}{553}.
%
\bibitem{smoot} Ivone F.M. Albuquerque and George F. Smoot hep-ph/0102078.
%
\bibitem{lipari} P.Lipari, M.Lusignoli and F. Sartogo,
\prl{74}{1995}{4384}; \\ P.Lipari, M.Lusignoli
\pr{D57}{1998}{3842}
%
\bibitem{sno} Q. Ahmad et al., SNO Collaboration, nucl-ex/0106015.

\end{thebibliography}
\end{document}